\documentclass[aps,prd,twocolumn,groupedaddress,showpacs]{revtex4}
\usepackage{amsfonts,amssymb,amscd,amsmath}
\usepackage{graphicx}
\usepackage{epsfig}
\usepackage{latexsym}
\usepackage{amssymb}
\usepackage{setspace}

\begin{document}

\def\abs#1{ \left| #1 \right| }
\def\lg#1{ | #1 \rangle }
\def\rg#1{ \langle #1 | }
\def\lrg#1#2#3{ \langle #1 | #2 | #3 \rangle }
\def\lr#1#2{ \langle #1 | #2 \rangle }
\def\me#1{ \langle #1 \rangle }

\newcommand{\bra}[1]{\left\langle #1 \right\vert}
\newcommand{\ket}[1]{\left\vert #1 \right\rangle}
\newcommand{\bx}{\begin{matrix}}
\newcommand{\ex}{\end{matrix}}
\newcommand{\be}{\begin{eqnarray}}
\newcommand{\ee}{\end{eqnarray}}
\newcommand{\nn}{\nonumber \\}
\newcommand{\no}{\nonumber}
\newcommand{\de}{\delta}
\newcommand{\lt}{\left\{}
\newcommand{\rt}{\right\}}
\newcommand{\lx}{\left(}
\newcommand{\rx}{\right)}
\newcommand{\lz}{\left[}
\newcommand{\rz}{\right]}
\newcommand{\p}{\partial}
\newcommand{\ts}{\times}
\newcommand{\ld}{\lambda}
\newcommand{\al}{\alpha}
\newcommand{\bt}{\beta}
\newcommand{\ga}{\gamma}
\newcommand{\si}{\sigma}
\newcommand{\bb}{\mathcal}
\newcommand{\dg}{\dagger}
\newcommand{\og}{\omega}
\newcommand{\Ld}{\Lambda}
\newcommand{\m}{\mathrm}
\newcommand{\vp}{\varrho}
\newcommand{\Og}{\Omega}

\title{The dynamical role of initial correlation
in the exactly solvable dephasing model}
\author{Yang Gao}
\email{gaoyangchang@gmail.com} \affiliation{Department of Physics,
Xinyang  Normal University, Xinyang, Henan 464000, China}

\begin{abstract}
We investigate the effects of the initial correlation on the
dynamics of open system in the exactly solvable pure dephasing
model. We show that the role of the initial correlation come into
play through a phase function and a weight factor, which would
perform oscillations during time evolution, and find that the
decoherence of a qubit coupled to a boson bath is more enhanced with
respect to a spin bath in the short time. We also demonstrate that
the trace distance between two states of a qubit can increase above
its initial value, and that the initial correlation can provide
another resource for the damply oscillation and revival of the
entanglement of two qubits. We finally investigate the dependence of
the crossover of decoherence from the dynamical enhancement to
suppression under the bang-bang pulse control on the initial
correlation and the statistics of the bath constituents.
\end{abstract}
\maketitle

\section{Introduction}

The conventional method of studying open quantum system is the
quantum master equation in the Lindblad form, which relies on two
assumptions \cite{open}. The first is that the coupling of the
system-bath interaction is weak, i.e. Born-approximation, and the
other is that the relaxation time of bath is much shorter than the
response time of system, i.e. Markov-approximation. Then the second
order perturbation is applied and the memory effect of the bath can
be neglected for simplification. If we ignore these assumptions, as
for the rigorous treatment of a quantum Brownian particle
interacting with a heat bath \cite{hpz}, a generalized exact master
equation or Hu-Paz-Zhang equation is derived, which does not have
Lindblad form. For general open systems, it is usually impossible to
solve such exact master equations analytically.

Besides these two assumptions, there is another underlying
assumption in deriving most of master equations in literature
\cite{open,hpz}, namely the open system is completely isolated from
bath initially. This uncorrelated initial condition is much useful
when the system-bath coupling is sufficiently weak and when we are
only interested in the long time behavior of system. In such cases,
the evolution of the system can be simply described by the
completely positive map \cite{cpm} acting only on the system state,
which leads to many profound results, e.g. the trace distance
between two states of the open system can not increase above its
initial value \cite{tot,measure,distance}. Although the uncorrelated
initial condition is widely used, it is mainly based on mathematical
idealization or simplicity rather than physical considerations, and
thus not always well justified when the system-bath coupling is
strong and when we study the transient behavior of system at short
times \cite{cpm,trans}.

Thus the effects of initial correlations on the dynamics of open
systems have been intensively discussed recently
\cite{tot,measure,distance}. It is argued in that the initial
correlation can be witnessed by the increase of the trace distance
of two system states over its initial value. Some other witnesses
for initial correlations have been proposed in \cite{purity}, such
as purity, quantum discord, entanglement, etc. Moreover, it has been
discussed that the initial condition can significantly influence the
time development of system based on some solvable models
\cite{dephase}. In the pure dephasing model with the boson bath and
the correlated initial condition, the decoherence function can have
some sharp peaks in comparison with the uncorrelated initial
condition \cite{trans,initial}.

In this paper we address the influence of the initial correlation on
the dynamics of the open system in the exactly solvable pure
dephasing model. We show that the role of the initial correlation
comes into play through a phase function and a weight factor, which
take into account the memory effect of bath and might give rise to
coherence oscillation during the time evolution. In particular, for
the short time, the decoherence of a qubit coupled to a boson bath
is more enhanced with respect to a spin bath. We also demonstrate
that the trace distance between two states of a qubit can increase
above its initial value, and that the entanglement of two qubits,
each locally interacting with an independent bath, can damply
oscillate and revive to a large amount being comparable to the
uncorrelated initial condition \cite{ent}. Moreover, it is known
that the decoherence can be effectively inhibited with the bang-bang
control pulses \cite{dd} and depends on the statistics of the bath
constituents \cite{spinc} without considering initial correlations.
So we finally extent them and investigate how the initial
correlations affect the decoherence and its crossover from the
dynamical enhancement to suppression under the pulse control.

The organization of this paper is as follows. In Sec. II we review
the main features of two pure dephasing models with boson and spin
baths respectively. The dynamics without initial correlations are
analyzed in Sec. III, including the bang-bang control pulses. We
then study the influence of the initial correlation on the dynamics
of the qubit in Sec. IV. The numerical results are given in Sec. V.
Our conclusion is given in Sec.VI.

\section{Models for pure dephasing}

The Hamiltonian describing a two-state system ($S$), i.e. qubit
coupled to a bath ($B$) for pure qubit-dephasing in the boson-bath
model can be written as \cite{dephase,dd,model} \be H &=&
H_S+H_B+H_\m{int}\nn &=& {\og_0} S_z +\sum_k \og_k a_k^\dg a_k+ S_z
\sum_k g_k( a_k^\dg + a_k ), \label{bm} \ee where $\og_0$ is the
excited energy of the qubit, and $S_z$ is the spin matrix of $z$
component with basis $2S_z \lg \pm = \pm \lg\pm$. The annihilation
and creation operators $a_k$ and $a_k^\dg$ are the bath mode with
frequency $\og_k$. The coupling strength between the qubit and the
$k$th bath mode is denoted by $g_k$. Suppose the initial state of
the system at $t=0$ is given by a density matrix $\rho(0)$, then the
state at time $t$ is \be \rho(t)=e^{-i H t} \rho(0) e^{i H t}. \ee
The reduced density matrix of the qubit is the partial trace taken
over the bath modes of the total density matrix,
$\rho_S=\m{Tr}_B\rho$.

It would be convenient to work in the interaction picture, where the
new state and time evolution are given by \be \vp(t) &=&
e^{i(H_S+H_B)t}\rho(t)e^{-i(H_S+H_B)t}, \nn \bb U(t,t_0) &=& \bb T
\exp[-i \int_{t_0}^tds \bb H(s)], \nn \bb H(t) &=& S_z\sum_k
g_k(e^{i\og_k t} a_k^\dg + e^{-i\og_k t}a_k). \ee Here the notation
$\bb T$ represents the time-ordered product. The evolution of the
new density matrix can be written as \be \vp(t)= \bb U(t,t_0)
\vp(t_0) \bb U^\dg(t,t_0). \label{time} \ee In order to further
simplify $\bb U(t,t_0)$, we use the generalized Baker-Hausdorf
formula \cite{initial}, \be && \bb T \exp \lz i \int_{t_0}^t ds A(s)
\rz = \exp \lz i \int_{t_0}^t ds A(s) \rz \nn && \indent \times \exp
\lt-{1 \over 2} \int_{t_0}^t ds_1\int_{t_0}^{s_1} ds_2
[A(s_1),A(s_2)] \rt, \label{baker} \ee which is satisfied if the
commutator $[A(s_1),A(s_2)]$ is a $c-$number. For the interaction
Hamiltonian $\bb H(s)$ we have \be [\bb H(t),\bb H(t')] &=&
-\frac{i}{2} \sum_k g_k^2\sin\og_k(t-t'). \label{comm} \ee With Eqs.
(\ref{baker}) and (\ref{comm}), the evolution operator $\bb
U(t,t_0)$ can be simplified as \be \bb U(t,t_0) &=& U_+(t,t_0) \lg
+\rg++U_-(t,t_0)\lg- \rg- , \label{boson} \ee where the unitary
operators $U_\pm$ are \be U_\pm &=& \exp \bigg \{ -if(t-t_0) \pm {1
\over 2}\sum_k \big [ \xi_k(t-t_0) e^{i\og_kt_0} a_k^\dg \nn && -
\xi_k^*(t-t_0) e^{-i\og_kt_0}a_k \big ] \bigg \}. \label{uni} \ee
The functions $f(t)$ and $\xi_k(t)$ are determined by the coupling
to the bath, \be f(t) &=& \sum_k {g_k^2 \over 4\og_k^2}(\og_k
t-\sin\og_k t), \nn \xi_k(t) &=& {g_k \over \og_k}(1-e^{i \og
t}).\ee

Similarly, the Hamiltonian for a qubit coupled to a spin-bath is
given by \cite{spinc} \be H &=& H_S+H_B+H_\m{int} \nn &=& {\og_0}
S_z +\sum_k \og_k \frac{\si_k^z}{2}+ S_z \sum_k g_k( \si_k^+ +
\si_k^- ), \label{sha} \ee where the coupling strength $g_k$ is
identical to that of Eq. (\ref{bm}), and $\og_k$ denotes the excited
energy of the $k$th bath spin $\vec{\si}_k$. It is worth to point
out that we have introduced an extra factor of $1/2$ in $H_B$ that
is different from the corresponding Hamiltonian in \cite{spinc} to
make the excited energy same as the photon bath, which leads to the
same evolutions for qubit in the two baths with zero temperature at
weak coupling regime and $t \to 0$ as shown below. This spin-bath
model can be solved by expressing the Hamiltonian $H$ as \be H &=&
H^+ \lg + \rg + + H^- \lg - \rg - , \nn H^\pm &=& \pm {\og_0 \over
2} + \frac{1}{2}\sum_k (\og_k \si_k^z \pm g_k \si_k^x). \ee The
unitary evolution operator $U(t)=e^{-i H t}$ is given by \be U(t)
&=& U_+ \lg + \rg + + U_- \lg - \rg - , \nn U_\pm &=& e^{\mp i\og_0
t / 2} \prod_k U_k^\pm , \label{spin} \ee where the unitary
$U_k^\pm$ is \be U_k^\pm &=& \cos \frac{\Og_k t}{2}-i\lx
\frac{\og_k}{\Og_k} \si_k^z \pm \frac{g_k}{\Og_k} \si_k^x
\rx\sin\frac{\Og_k t}{2}, \nn \Og_k &=&\sqrt{\og_k^2+g_k^2}.
\label{spu} \ee

It can seen from Eqs. (\ref{boson}) and (\ref{spin}) that the
populations of the $\lg \pm$ states, $\rho_S^{++}(t)$ and
$\rho_S^{--}(t)$ do not change with time, namely the pure dephasing
model can not describe the relaxation process of the qubit to
equilibrium. However, the exact solutions of these simple models
with initial correlations still capture the essential features of
decoherence in the more complicated dissipative models.

\section{Dephasing without initial correlations}

The uncorrelated initial condition means that the initial state of
the total system is a direct product of the qubit state and the bath
state, \be \rho(0) &=& \rho_S(0)\otimes \rho_B \nn \rho_S(t) & = &
\mathrm{Tr}_B[U \rho_S(0)\otimes \rho_B U^\dg]. \label{uncon} \ee
Note the map from $\rho_S(0)$ to $\rho_S(t)$ is completely positive
under the uncorrelated initial condition. Usually, the initial bath
is assumed to be in the thermal equilibrium at temperature
$T=1/\bt$, $\rho_B=e^{-\bt H_B}/Z_B$ with the normalization
$Z_B=\mathrm{Tr}_B[e^{-\bt H_B}]$. Because the evolution is governed
by the unitary operator of the form $U_+ \lg + \rg + + U_- \lg - \rg
-$, the coherence of the qubit $\rho_S^{+-}$ evolves in time as \be
\rho_S^{+-}(t) &=& e^{-i\og_0 t}e^{-\Gamma_a(t)}\rho_S^{+-}(0), \nn
\Gamma_a(t) &=& -\ln \lx \m {Tr}_B [U_+ \rho_B U_-^\dg] \rx \nn &=&
-\ln \lx \m {Tr}_B [\rho_B U_-^\dg U_+] \rx, \label{diff} \ee where
$\Gamma_a(t)$ denotes the decoherence function for the boson-bath
model. Using Eq. (\ref{uni}) and the identity \be && \m {Tr}_B
[\rho_B \exp \lx \sum_k \eta_k a_k^\dg-\eta_k^* a_k \rx] \nn &&
\indent = \exp \lx -\sum_k {|\eta_k|^2 \over 2}\coth{\bt \og_k \over
2} \rx, \label{bs0} \ee we obtain \be U_-^\dg U_+ &=& \exp \bigg \{
\sum_k \big [ \xi_k(t) a_k^\dg - \xi_k^*(t) a_k \big ] \bigg \}, \nn
\Gamma_a(t) &=& \sum_k g_k^2\coth{\bt \omega_k \over 2}{ 1-\cos
\og_k t \over \og_k^2}. \label{bs} \ee

For the spin-bath model under the uncorrelated initial condition, we
use Eqs. (\ref{spin}), (\ref{spu}), and (\ref{uncon}) to get \be
\rho_S^{+-}(t) &=& e^{-i\og_0 t}e^{-\Gamma_\sigma(t)}\rho_S^{+-}(0),
\nn \Gamma_\sigma(t) &=& - \ln \bigg \{ \m {Tr}_B [ \rho_B \prod_k
(U_k^-)^\dg U_k^+] \bigg \} \nn &=& -\sum_k \ln\lx
1-2\frac{g^2_k}{\Og_k^2}\sin^2 \frac{\Og_k t}{2} \rx . \label{sp}
\ee Eqs. (\ref{bs}) and (\ref{sp}) are the exact results for the
decoherence functions under the uncorrelated initial conditions in
the pure dephasing models. Unlike the boson-bath model, the
decoherence function for the spin-bath model is independent of
temperature and can lead to complete dephasing when
$1-2(g^2_k/\Og_k^2)\sin^2 ({\Og_k t/2})=0$ at certain time if one of
the coupling strength $g_k\ge \og_k$. This is quite different from
Eq. (\ref{bs}) in the boson-bath model, which does not have complete
dephasing even if all the couplings satisfy $g_k\ge \og_k$. If the
size of bath is small, the coherence can revive to a large amount at
certain time. Moreover, due to the renormalized spin-bath mode
frequency $\Og_k>\og_k$, it induces faster dynamics than boson-bath
at zero temperature despite the same bath coupling spetrum.

In the short time case $t \to 0$, we see that both Eqs. (\ref{bs})
and (\ref{sp}) perform the quadratic behavior, namely $\Gamma(t)
\propto t^2$. On the other hand, in the weak coupling limit $g_k \ll
\og_k$ and $\Omega_k \approx \og_k$, we can see Eq. (\ref{sp})
approaches to (\ref{bs}) at zero temperature since $-\ln (1-x)
\approx x$ for $x \ll 1$. It is known that for the boson-bath model
the non-Markovian master equation up to the second order
perturbation can give the exact result for the decoherence function
even in the strong coupling limit because the expansion of the
evolution unitary operator can be truncated at the second term,
which is not legitimate for the spin-bath model. The quantum master
equation in the second-order approximation is \cite{open,second} \be
\frac{d \vp_S(t)}{d t}=-\int _0^t ds ~ \m {Tr}_B [\bb H(t),[\bb
H(s),\vp_S(t)\otimes \vp_B]]. \ee For the pure dephasing model, it
leads to \be \frac{d \vp_S^{+-}(t)}{d t}=-\int _0^t ds \bb K(t,s)
\vp_S^{+-}(t), \label{mas} \ee where the kernel for the boson bath
is given by \be \bb K(t,s)=\sum_k g_k^2 \cos[\og_k (t-s)] \coth
\frac{\bt \og_k}{2}. \ee The solution of Eq. (\ref{mas}) gives the
decoherence function $\Gamma^{ME}_a(t)$ of the form (\ref{bs}),
which is valid as long as the second order analysis is legitimate,
i.e. $g_k/\og_k \leq 1$. On the other hand, in the spin-bath model
the kernel is \be \bb K(t,s)=\sum_k g_k^2 \cos[\og_k (t-s)].\ee The
solution of Eq. (\ref{mas}) is then \be \Gamma_\sigma^{ME}(t) =
\sum_k \frac{g_k^2}{\og_k^2} (1-\cos\og_k t), \ee which is also
independent of temperature and approaches to (\ref{sp}) only when
$g_k/\og_k \ll 1$. Therefore, the condition of the validity of the
second order master equation for the spin-bath is more stringent
than that of the boson-bath. We also note that coherence dynamics
described by Eqs. (\ref{bs}) and (\ref{sp}) in both models are
non-divisible and non-Markovian according to definition in
\cite{dm}.

Next, we analyze the effect of the dynamical decoupling pulses along
$x$ direction at intervals $\tau$ under the uncorrelated initial
condition \cite{dd,spinc}. The simplest bang-bang pulses that can
significantly reduce the dephasing rate is realized by frequent
$\pi$ pulses along $x$ direction applied on the qubit. That is after
such a pulse, the qubit states change as $\lg \pm \to \lg \mp$ in
the Schr\"odinger picture. In the presence of the decoupling pulses,
at time $t=2N\tau$, the evolution unitary operator becomes \be \bb
U(t,0) = \bb U_+ \lg + \rg + + \bb U_- \lg - \rg -, \label{seq} \ee
where \be \bb U_\pm = e^{\pm i\og_0 t/2} U_\mp(t,t-\tau)\cdots
U_\mp(2\tau,\tau)U_\pm(\tau,0), \ee and the phase factors $e^{\pm
i\og_0 t/2}$ appear due to the state transformation from the
Sch\"odinger picture to the interaction picture. Substituting Eqs.
(\ref{uni}) into (\ref{seq}) and neglecting state-independent global
phase factors that are irrelevant to the density matrix, we can
treat the factors in Eq. (\ref{seq}) as commuting operators and get
\be \bb U_\pm = e^{\pm i\og_0 t/2} \exp \lt \pm {1 \over 2} \sum_k
[\eta_k(\tau) a_k^\dg - \eta_k^*(\tau)a_k ]\rt, \label{newseq} \ee
where \be \eta_k(\tau)= \xi_k(\tau)(1-e^{i\og_k \tau})\sum_{n=1}^N
e^{2i(n-1)\og_k \tau}.\ee Repeat the steps to derive Eq. (\ref{bs}),
we have \be \rho_S^{+-}(t) &=& e^{-\Gamma_a^\pi(t)}\rho_S^{+-}(0),
\label{bsu} \\ \Gamma_a^\pi(t) &=& \sum_k g_k^2\coth{\bt \omega_k
\over 2}\tan^2{\omega_k \tau \over 2}{ 1-\cos \og_k t \over
\omega_k^2}. \no \ee

In the spin-bath model, the evolution under periodic $\pi$ pulses is
given by the unitary \be U(t)=(U_-U_+)^N \lg + \rg + + (U_+U_-)^N
\lg - \rg -. \ee Hence it is sufficient to find a closed form for
the matrix $(U_-U_+)^N$. Since \be (U_-U_+)^N = \prod_k
[U_k^-U_k^+]^N \ee and \be U_k^-U_k^+ &=& x_k I - i \lz y_k
\si_k^z+\sqrt{1-x_k^2-y_k^2} ~ \si_k^y \rz , \\  x_k &=& 1-2{\og_k^2
\over \Og_k^2}\sin^2 \frac{\Og_k \tau}{2}, \qquad\ \ y_k =
\frac{\og_k}{\Og_k} \sin \Og_k \tau, \no \ee the expression of
$[U_k^-U_k^+]^N $ can be evaluated by writing $U_k^-U_k^+$ in terms
of its eigenvectors. That is if \be U_k^-U_k^+ = \ld_k^+ \lg {v_k^+}
\rg {v_k^+} + \ld_k^- \lg {v_k^-} \rg {v_k^-}, \ee we have \be
(U_k^-U_k^+)^N = (\ld_k^+)^N \lg {v_k^+} \rg {v_k^+} + (\ld_k^-)^N
\lg {v_k^-} \rg {v_k^-}\ee where the corresponding eigenvalues and
eigenvectors are given by \be \ld_k^\pm &=& x_k\pm i\sqrt{1-x_k^2}
\nn \lg {v_k^{\pm}} &=&\frac{1}{\sqrt{1+\al_\pm^2}}[\lg +_k-i
\al_\pm \lg -_k],\nn \al_\pm &=& \frac{y_k \pm
\sqrt{1-x_k^2}}{\sqrt{1-x_k^2-y_k^2}}. \ee The eigenvalues of
$U_k^+U_k^-$ are the same as $U_k^-U_k^+$, and the corresponding
eigenvectors are given by the replacements $\al_\pm \to -\al_\pm$.
The final result we obtained is \be \rho_S^{+-}(t) &=&
e^{-\Gamma_\sigma^\pi(t)}\rho_S^{+-}(0), \label{fsu} \\
\Gamma_\sigma^\pi (t) &=& -\sum_k \ln \lx 1- 8 F_k^2 \rx ,\nn F_k
&=& \lx{\sin N \phi_k \over \sin \phi_k}\rx {g_k \og_k \over
\Og_k^2} \sin^2 \frac{\Og_k \tau}{2}, \no \ee where $\phi_k=
\cos^{-1} x_k$. Under the periodic pulse control, the expression for
$\Gamma_\sigma^\pi(t)$ is still temperature-independent and
approaches to $\Gamma_a^\pi(t)$ in (\ref{bsu}) at zero temperature
as in the weak coupling limit $g_k/\og_k \ll 1$ and $\phi_k \approx
\og_k \tau$, which can also be obtained from the second order
analysis.

\section{Dephasing with initial correlations}

One natural way to implement the initial correlation between the
qubit and bath is to use the positive operator-value measurement
(POVM) \cite{open} $E_m$ acting only on the qubit, where the whole
system is in the thermal equilibrium state $\rho=e^{-\bt H}/\m
{Tr}[e^{-\bt H}]$, then after the action of POVM, the whole system
becomes to be \be \rho(0)=\frac{1}{Z} \sum_m E_m e^{-\bt H} E_m^\dg,
\label{con} \ee where the factor $Z$ is the normalization of
$\rho(0)$, \be Z &=& \sum_{\pm} u_\pm \m {Tr}_B [e^{-\bt H_\pm}], \\
H^{\pm} &=& \pm {\og_0\over 2} +\sum_k \og_k a_k^\dg a_k \pm
\frac{1}{2}\sum_k g_k( a_k^\dg + a_k ), \no \ee where the notation
$u_\pm \equiv \sum_m \lrg \pm {E_m^\dg E_m} \pm$, and the fact that
the total Hamiltonian is diagonal in the basis of $S_z$ has been
used.

In such way we prepare an initial state by measurement, instead of
the supposed uncorrelated state Eq. (\ref{uncon}). In contrast tothe
form of (\ref{uncon}), the initial density matrix (\ref{con}) is
expressed in terms of the total Hamiltonian $H$ and takes into
account the initial qubit-bath correlation through the interaction
term $H_{\m{int}}$ of $H$. Consequently, the bath is no longer in
thermal equilibrium initially, and its initial state becomes \be
\rho_B(0) &=& \m {Tr}_S[\rho(0)] = \frac{1}{Z} \sum_m \m {Tr}_S [E_m
e^{-\bt H} E_m^\dg] \nn & = & \frac{1}{Z} \sum_{\pm} u_\pm e^{-\bt
H^{\pm}}. \label{init} \ee  Eq. (\ref{init}) is quite different from
the thermal state $\rho_B=e^{-\bt H_B}/Z_B$ which does not contain
the interaction terms. Even the resulting state sometimes takes the
initial product form $\rho_S \otimes \rho_B(\rho_S)$ where $\rho_B$
depends on $\rho_S$, the dynamical map would be very different from
the situation of $\rho_B$ and $\rho_S$ being purely independent. For
the purely independent case, the map from $\rho_S(0)$ to $\rho_S(t)$
is linear, whereas the map for the former case is nonlinear.
Therefore, in the following we will not distinguish the usual
classification of quantum and classical correlations. For example in
\cite{class} it was shown that even the classical correlation can
lead to entanglement oscillation. So the product states $\rho_S
\otimes \rho_B(\rho_S)$ are also correlated in our general sense.

Now we calculate the evolution of coherence $\rho_S^{+-}(t)$ with
the initial correlation introduced by Eq. (\ref{con}). Using Eq.
(\ref{time}), we first write \be \vp_S^{+-}(t) &=& \m {Tr} [\si_-
\vp(t)] \nn & = & \frac{1}{Z} \sum_m \m {Tr}_B \m {Tr}_S [\si_- \bb
U(t) E_m e^{-\bt H} E_m^\dg \bb U^\dg(t)] \nn &=& \frac{1}{Z}
\sum_{m,\pm} \lrg \pm {E_m^\dg \si_- E_m} \pm \m {Tr}_B [U_-^\dg U_+
e^{-\bt H_\pm}] \nn &=& \frac{1}{Z} \sum_{\pm} w_\pm \m {Tr}_B
[U_-^\dg U_+ e^{-\bt H_\pm}], \ee where the factor $w_\pm$ is $
w_\pm = \sum_m\lrg \pm {E_m^\dg \si_- E_m} \pm$. As $\m {Tr}_B
[U_-^\dg U_+ e^{-\bt H_B}]$ appeared in Eq. (\ref{diff}), we need to
find the expression for $\m {Tr}_B [U_-^\dg U_+ e^{-\bt H_\pm}]$.
Note the identity \be && {1 \over Z_B} \m {Tr}_B [e^{-\bt H_\pm}
e^{\eta_k a_k^\dg-\eta_k^* a_k}] = \m {Tr}_B [\rho_B e^{\eta_k
a_k^\dg-\eta_k^* a_k}] \nn && \indent \times \exp\bigg[ \bt
\sum_k{g_k^2 \over 4\og_k} \mp {\bt \og_0 \over 2} \pm i\Theta
\bigg], \ee where a unitary transformation that diagnoses $H^\pm$
was performed inside the trace function, and the phase is \be \Theta
= \sum_k {g_k \over 2 i \og_k} (\eta_k^* -\eta_k). \ee Using Eqs.
(\ref{bs0}) and (\ref{bs}), we get the final result for the
coherence \be \rho_S^{+-}(t) = \rho_S^{+-}(0)e^{-\Gamma_a(t)}
\frac{\sum_\pm w_\pm e^{\mp \bt \og_0/2}e^{ \pm
i\Theta_a(t)}}{\sum_\pm w_\pm e^{\mp \bt \og_0/2}}\nn =
\rho_S^{+-}(0)e^{-\Gamma_a(t)} [ \cos \Theta_a(t)+i W
\sin\Theta_a(t) ], \label{bsc} \ee where the decoherence function
$\Gamma_a(t)$ has the same form of (\ref{bs}). The phase function is
given by \be \Theta_a(t) = \sum_k g_k^2 \frac{\sin\og_k t}{\og_k^2},
\ee and the weight factor is \be W &=& \frac{w_+ e^{- \bt
\og_0/2}-w_- e^{\bt \og_0/2}}{w_+ e^{-\bt \og_0/2}+w_- e^{\bt
\og_0/2}}, \ee which take into account the memory effect of bath and
the role of initial correlation. Here the initial coherence is given
by \be \rho_S^{+-}(0) &=& \frac{w_+ e^{-\bt \og_0/2}+w_- e^{\bt
\og_0/2}}{u_+ e^{-\bt \og_0/2}+u_- e^{\bt \og_0/2}}. \ee Similar
results have been obtained in \cite{initial}.

For the spin-bath model, we get \be \rho_S^{+-}(t) &=&
\rho_S^{+-}(0)e^{-\Gamma_\sigma(t)} [\cos \Theta_\sigma(t)+i W
\sin\Theta_\sigma(t)] \nn && \times \prod_k \sqrt{1+\eta_k^2(t)},
\label{fc} \ee where \be \Theta_\sigma(t) &=& \sum_k \tan^{-1}
\eta_k(t), \nn \eta_k(t) &=& { \frac{g^2_k}{\Og_k^2} \tanh{\bt \Og_k
\over 2}{\sin \Og_k t} \over \lx 1-2\frac{g^2_k}{\Og_k^2}\sin^2
\frac{\Og_k t}{2} \rx }. \ee The memory effect of bath is taken into
account by the temperature dependent function $\eta_k(t)$, which
also makes the decoherence function temperature independent. For
convenience, we express Eqs. (\ref{bsc}) and (\ref{fc}) in a compact
form, \be \rho_S^{+-}(t) &=& \rho_S^{+-}(0)e^{-\Gamma(t)} [\cos
\Theta(t)+i W \sin\Theta(t)] \nn && \times \prod_k \lz 1+\eta_k^2(t)
\rz ^{s/2}, \ee where the exponent $s=0$ for the boson bath and
$s=1$ for the spin bath. At $T=0$ and $g_k \to 0$, $\eta_k \to
g_k^2\sin\og_k t/\og_k^2$ and $\Theta_\sigma \to \Theta_a$. In such
regime, the spin bath would give a bit larger result than the photon
bath for coherence function due to the extra $\sqrt{1+\eta_k^2(t)}$.

Next, we consider the quantification of the degree of initial
correlation \cite{measure}. We note that the trace distance of any
two states $\rho_1$ and $\rho_2$ defined by $D(\rho_1,\rho_2)\equiv
|\rho_1-\rho_2|/2$ with $|A|=\mathrm{Tr} \sqrt{A^\dg A}$ cannot
increase above its initial value under the completely positive map,
and the dynamical map from $\rho_S(0)=\mathrm{Tr}_B[\rho(0)] $ to
$\rho_S(t)= \mathrm{Tr}_B[U \rho(0) U^\dg]$ is completely positive
if $\rho(0)$ is in a product state \cite{cpm}. So a witness for
initial system-bath correlation is proposed in \cite{measure},
namely the amount of trace distance of two states increased above
its initial value. For example, we consider the quantity
$D(\rho_S(t),\bar{\rho}_S(t))$, where $\bar \rho_S(t)$ is
constructed from the marginal state of $\rho(0)$, i.e. $\bar
\rho_S(t) \equiv \mathrm{Tr}_B [U \bar{\rho}(0) U^\dg] $ with
$\bar{\rho}(0) = \rho_S(0) \otimes \rho_B(0)$. We have obviously
$D(\rho_S(0),\bar{\rho}_S(0))=0$, so it can be used as a witness for
the initial correlation as argued in \cite{measure} and an upper
bounded was also found to be $D(\rho(0),\bar{\rho}(0))\ge
D(\rho_S(t),\bar{\rho}_S(t))$. By direct calculations, we obtain \be
D(\rho_S(t),\bar{\rho}_S(t)) &=& |\rho_S^{+-}(0)
(W-W_u)\sin\Theta|e^{-\Gamma} \nn && \times \prod_k \lx 1+\eta_k^2
\rx ^{s/2} , \ee where \be W_u &=& \frac{u_+ e^{- \bt \og_0/2}-u_-
e^{\bt \og_0/2}}{u_+ e^{-\bt \og_0/2}+u_- e^{\bt \og_0/2}}.\ee If
$W-W_u \neq 0$, the trace distance can take nonzero value as time
going on, which indicates the initial correlation between the qubit
and bath.

However, the above witness does not cover the case of the initial
correlated product state $\rho(0)=\rho_S \otimes \rho_B(\rho_S)$. To
illustrate the dynamical role of such type of initial correlation,
we find that it can provide a new resource for the entanglement
oscillation and revival. Suppose two qubits each locally interacting
with an independent bath and the initial state is prepared by the
operator $E_m =\lg \Psi \rg \Psi \otimes I_B$ with $\lg \Psi=(\lg
{++} + \lg {--})/\sqrt{2}$, unlike the case of uncorrelated initial
condition, which induces initial entanglement between the two
independent baths through the correlation of qubit-bath, and gives
an "X" structure density matrix maintained during the evolution
\cite{ent}. The entanglement measure of the two qubits, i.e. the
Wootters concurrence, can then be obtained by \cite{wootters} \be
C(t)=\max \lz 0,2|\rho_{14}(t)|-2\sqrt{\rho_{22}(t)\rho_{33}(t)}
\rz, \ee where the standard product basis $\{ \lg 1 \equiv \lg {++},
\lg 2 \equiv \lg {+-},\lg 3 \equiv \lg {-+},\lg 4 \equiv \lg {--}\}$
were used and the nonzero element $\rho_{14}(0)=1/2$. For such
initial state, we have $\rho_{22}(t)=\rho_{33}(t)=0$, and \be
\rho_{14}(t) &=& \rho_{14}(0) e^{-2\Gamma}(\cos 2\Theta+i W_e \sin
2\Theta)\nn && \times \prod_k (1+\eta_k^2)^s, \ee where the weight
factor is $W_e=-\tanh \bt \og_0 <1$. The result for the uncorrelated
initial condition is obtained by putting $\Theta=\eta_k=0$ in the
above equations. If we replace the initial state by $\lg \Phi=(\lg
{+-} + \lg {-+})/\sqrt 2$, the effect of initial correlation would
cancel out in the phase function, namely $\Theta=0$ in the
entanglement measure.

The dynamical decoupling effect of the periodic pulses can be
derived as in the previous section. The final results have the same
forms of (\ref{bsc}) and (\ref{fc}) with the replacements $\Gamma
\to \Gamma^\pi$, $\Theta_a \to \Theta_a^\pi$, and
$\eta_k\to\eta_k^\pi$, where \be \Theta_a^\pi(t) &=& \sum_k
g_k^2 \tan {\og_k \tau \over 2}\frac{1-\cos\og_k t}{\og_k^2}, \label{bphase} \\
\eta_k^\pi(t) &=& {2 \frac{g_k^2}{\Og_k^2} \tanh{\bt \Og_k \over 2}
\tan \frac{\Og_k \tau}{2} \sin^2 N \phi_k \over \lx 1 +
\frac{g_k^2}{\Og_k^2}\tan^2 \frac{\Og_k \tau}{2}\cos 2N\phi_k \rx }.
\ee Obviously, in the limit of the pulse interval $\tau \to 0$, we
have $\Gamma^\pi \to 0$, $\Theta_a^\pi \to 0$ and $\eta_k^\pi \to
0$, which means that the decoherence are inhibited significantly.
For the finite pulse interval, the behaviors of the two models are
quite different as discussed below.

\section{Numerical result}

Because the decoherence function for the boson bath linearly depends
on the coupling strength $g_k^2$, we can treat the bath within the
classical noise description \cite{spinc,spina}. By introducing the
spectral density of the boson-bath, \be J(\og) &=& \sum_k g_k^2
\de(\og -\og_k ), \nn &\to & g^2 \sum_k \de(\og -\og_k ) \equiv g^2
\int_0^\infty d\og N(\og), \label{spect} \ee where the second line
holds for the all equal couplings with the boson bath, and the
number distribution of bath mode is denoted by $N(\og)$. Then the
summation over the bath modes in the decoherence functions can be
transformed into integral. For example, Eq. (\ref{bs}) becomes \be
\Gamma_a(t) &=& \int d\og J(\og) \coth{\bt \og \over 2} {1-\cos\og t
\over \og^2} \nn &\to & g^2 \int d\og N(\og) \coth{\bt \og \over 2}{
1-\cos \og t \over \og^2}. \label{bosn} \ee

For the spin bath, it can not be written as an integral as opposed
to the boson bath with the chosen bath spectrum since the mode
frequency is renormalized to be $\Omega_k=\sqrt{\og_k^2+g_k^2}$,
which nonlinearly depends on $g_k^2$ except in the weak coupling
regime \cite{spina}. In order to compare the two models with closed
forms, by inspecting Eq. (\ref{spect}) we suppose the qubit is
equally coupled with each mode of the bath, i.e. $g_k=g$ for all
$k$, but the number of each mode is distributed according to the
chosen spectrum $N(\og)$. For example, Eq. (\ref{sp}) can thus be
written as \be \Gamma_\sigma(t) &=& -\int d\og N(\og) \ln\lx
1-2\frac{g^2}{\Og^2}\sin^2 \frac{\Og t}{2} \rx , \label{zero} \ee
where the renormalized frequency is $\Og=\sqrt{\og^2+g^2}$. It needs
to point out that the above recipe only works well at the beginning
of the evolution for the spin bath model due to the possible zeroes
of the argument such as in Eq. (\ref{zero}). It can give large
coherence revival at large time even with a huge size bath which is
just an artifact of our particular assumption. This difficulty can
be avoid by randomly choosing coupling constants $g_k$, which can
give almost complete decoherence as the size of spin bath increases
\cite{zero}. Hence we only concern the beginning evolution of
coherence function in order to avoid the possible divergence
problems for spin bath in the following.

\begin{figure}[t!]
\begin{minipage}{.20\textwidth}
\centerline{\epsfxsize 50mm \epsffile{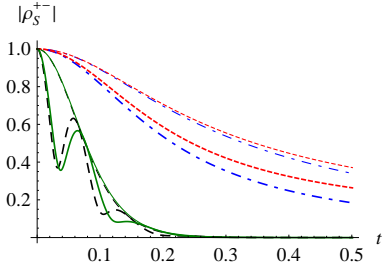}}
\end{minipage}
\caption{Time evolution of the coherence for the uncorrelated (thin
lines) and correlated (thick lines) initial conditions with the
Ohmic bath spectral density. The coherence in the weak (strong)
boson \& spin bath is represented by the dot-dashed (dashed) \&
dotted (solid) lines. Here the initial coherence is rescaled to
unity.}
\end{figure}

\begin{figure}[t!]
\begin{minipage}{.22\textwidth}
\centerline{\epsfxsize 42mm \epsffile{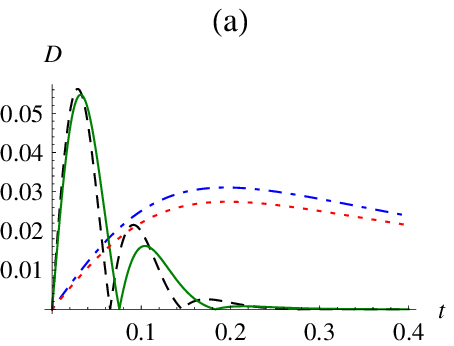}}
\end{minipage}
\begin{minipage}{.22\textwidth}
\centerline{\epsfxsize 42mm \epsffile{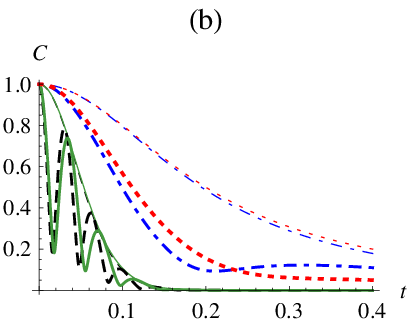}}
\end{minipage}
\caption{(a) The evolution of the trace distance of two states of
qubit. The dot-dashed (dashed) \& dotted (solid) line is for the
weak (strong) boson \& spin bath. (b) The evolution of the
concurrence of two qubits each locally interacting with an
independent bath. The concurrences with uncorrelated initial states
are marked by thin lines. \label{entangle}}
\end{figure}

\begin{figure}[t!]
\begin{minipage}{.19\textwidth}
\centerline{\epsfxsize 52mm \epsffile{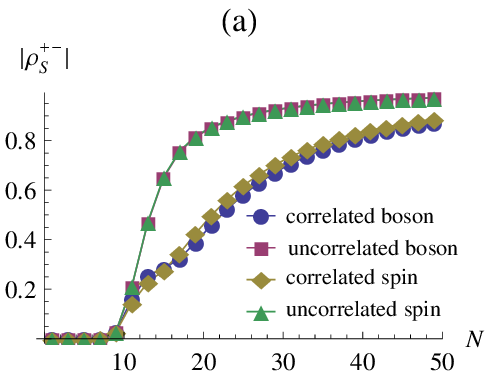}}
\end{minipage}
\begin{minipage}{.19\textwidth}
\centerline{\epsfxsize 42mm \epsffile{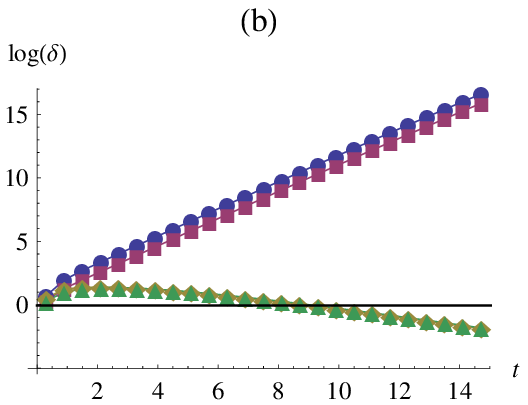}}
\end{minipage}
\caption{Time evolution of the coherence in the strong baths for the
uncorrelated and correlated initial conditions under the bang-bang
control pulses. (a) The dependence of the coherence at time $t=2$ in
the strong baths on the pulses applied at a rate $\tau = 1/N$. (b)
Time evolution of the relative change of the coherence
$\delta=|\rho_{S,\pi}^{+-}/\rho_S^{+-}|$ under pulse control applied
at a rate $\tau=0.15$ with respect to its free evolution in the weak
baths. }
\end{figure}

\begin{figure}[t!]
\begin{minipage}{.0\textwidth}
\centerline{\epsfxsize 60mm \epsffile{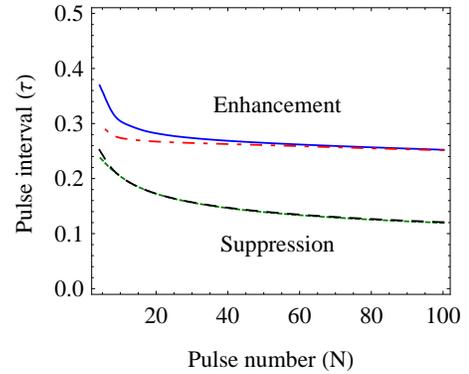}}
\end{minipage}
\caption{The crossover of the qubit coherence from the suppression
to the enhancement under the bang-bang control pulses is plotted as
a function of the pulse interval ($\tau$) versus the pulse number
($N$). The solid (dot-dashed) line is for the weak boson bath
with(out) initial correlation. The dashed (dotted) line is for the
weak spin bath with(out) initial correlation, where the dotted line
also represents the crossover in the zero temperature bath. For the
strong bath, the crossover merges into the case of the weak bath
without initial correlation.}
\end{figure}

For the numerical analysis, we choose the operator $E_m$ to be \be
E_m=I+{1 \over 2}(\si_x+\si_z), \ee and the constants $u_\pm$ and
$w_\pm$ are thus given by \be u_\pm = \frac{3}{2} \pm 1, \qquad
w_\pm = \frac{1}{2}\pm \frac{1}{4}.\ee Suppose the coupling constant
$g=0.02$, the temperature $\bt=2$, and the energy $\og_0=0.1$
\cite{limit}. We also set the number distribution of bath mode be
the Ohmic spectrum with a high frequency cut-off, $N(\og)=\ld N_0
\og e^{-\og/\Ld}$ with the cut-off $\Ld=5$ and $N_0=2.5\times
10^{3}$, where the weak \& strong bath corresponds to $\ld=1$ \&
$10$.

In Fig. 1 we show the free evolution of the coherence in the bath
with and without initial correlation. It can be seen that the
initial correlation can lead to damped coherence oscillation for the
strong bath, and reduce the coherence to some extent for the weak
bath. We also see that for the short time, the boson bath dephases
the qubit to more degree than the spin bath with the same initial
condition.

The evolution of the trace distance $D(\rho_S(t),\bar\rho_S(t))$ is
plotted in Fig. 2 (a), which can take nonzero value at $t>0$
witnessing the initial quantum correlation. Fig. 2 (b) shows that
the entanglement measure, i.e. concurrence $C$, can revive to a
large amount in contrast to the uncorrelated initial condition for
the strong bath, which only appears as the two qubits are coupled to
a common strong bath without initial correlation \cite{large}. It
can be seen that both quantities can perform damped oscillations for
the strong baths. The reason for the entanglement oscillation with
initial correlation is that the preparation scheme on the two qubits
induces some entanglement between the two independent baths at the
beginning, which can return into qubits at later time via the
non-Markovianity of the evolution \cite{ent}. Fig. 2 (b) also shows
that the entanglement without initial correlation always larger than
the envelop of the entanglement with initial correlation.

Fig. 3 (a) shows the dependence of the coherence at time $t=2$ in
the strong baths on the pulses applied at a rate $\tau = 1/N$, from
which we see that the initial correlation always reduces the
efficiency of the control pulses relative to the uncorrelated case.
Fig. 3 (b) shows the evolution of the relative change of the
coherence $\delta=|\rho_{S,\pi}^{+-}/\rho_S^{+-}|$ under pulse
control applied at a rate $\tau=0.15$ with respect to its free
evolution in the weak baths. It indicates that the $\pi$ pulse
control suppresses the decoherence for the boson bath, while the
same control can only suppress the decoherence for the spin bath in
the short time duration, and then crossovers to the enhancement of
the decoherence as time becomes longer. Moreover, the decoherence is
more strongly suppressed in the boson bath than in the spin bath.

In Fig. 4, we plot the crossover of the qubit coherence from the
suppression to the enhancement under the bang-bang control pulses as
a function of the pulse interval ($\tau$) versus the pulse number
($N$). It is shown that the initial correlation affects the
crossover of decoherence to some extent only at the beginning of
evolution, and induces relatively more changes for the boson (weak)
bath than for the spin (strong) bath. The underlying reason is that
the bath's memory of the initial correlation would be lost as time
being longer. On the other hand, the higher temperature for the
boson bath, the more efficient for the control pulses to make the
crossover occur. As shown in Fig. 4, the crossovers for the
correlated and uncorrelated initial conditions in the limit of $t\to
\infty$ are essentially coincident with each other. Asymptotically
for the boson bath, note that \be \frac{\Gamma_a(t)}{t} &=& 2g^2
\int_0^\infty d\og N(\og) \coth{\bt \og \over 2}{\sin^2\frac{\og
t}{2} \over \og^2 t} \nn &\to & \frac{\pi}{2}\kappa_\bt(0) \text {
when } t \to \infty, \ee and \be \frac{\Gamma_a^\pi(t)}{t} &=& 2g^2
\int_0^\infty d\og N(\og) \coth{\bt \og \over 2}\tan^2{\og \tau
\over 2}{\sin^2\frac{\og t}{2} \over \og^2t} \nn &\to &
\frac{4}{\pi} \kappa_\bt \lx \frac{\pi}{\tau} \rx \text { when } t
\to \infty \text { and } \tau \to 0, \ee where
$\kappa_\bt(\og)\equiv g^2 N(\og) \coth (\bt \og / 2)$ and the
following identities have been used \cite{limit}, \be && \lim_{t\to
\infty} t {\sin^2\frac{\og t}{2} \over \lx \frac{\og t}{2} \rx^2 } =
2\pi \delta(\og), \nn && \lim_{t\to \infty} t {\sin^2\frac{\og t}{2}
\over \lx \frac{\og t}{2} \rx^2 } \tan^2\frac{\og \tau}{2} = {8
\over \pi}\sum_{n=0}^\infty {1 \over (2n+1)^2} \nn && \indent \times
\lz \delta \lx \og+\frac{\pi}{\tau}(2n+1) \rx + \delta \lx
\og-\frac{\pi}{\tau}(2n+1) \rx \rz. \no \ee Here the pulse interval
$\tau$ is fixed and $t=2N\tau$ while the pulse number $N\to \infty$.
The crossover at distant time between the two regimes takes place at
$\tau=\tau^*$ where $\tau^*$ is determined by the equation, \be
\frac{4}{\pi} \kappa_\bt \lx \frac{\pi}{\tau^*} \rx
=\frac{\pi}{2}\kappa_\bt(0), \ee the solution of which is given by
$\tau^*=0$ at zero temperature and $\tau^*\approx 0.289$ at $\bt=2$
same as the numerical results shown in Fig. 4. For the spin bath,
the crossover under control shows the temperature insensitivity
instead.

\section{Conclusion}

We addressed the effects of initial correlation on the dynamics of
open system in the pure dephasing models with the boson and spin
baths. We found that the initial correlation can reveal the memory
of bath during the time evolution. The decoherence of a qubit
coupled to a boson bath is more enhanced with respect to a spin bath
in the short time. We also demonstrated that the trace distance
between two states of qubit can increase above its initial value
witnessing the initial non-classical correlation between the qubit
and bath, and that the entanglement of two qubits, locally
interacting with an independent bath, can damply oscillate and
revive to a large amount compared to the uncorrelated initial
condition. We finally showed that the initial correlation affects
the crossover of decoherence from the dynamical enhancement to
suppression under the bang-bang control pulses to some extent only
at the beginning of evolution, and induces more changes for the
boson (weak) bath than for the spin (strong) bath. On the other
hand, the higher temperature for the boson bath, the more efficient
for the control pulses to make the crossover occur. For the spin
bath, the crossover under control shows the temperature
insensitivity instead.

\begin{acknowledgments}
The authors would like to acknowledge support from NSFC grand No.
11147137.
\end{acknowledgments}


\end{document}